\documentclass[twocolumn]{emulateapj}
\newcommand{\etal}{et al.}
\newcommand{\photo}{{\tt photo}}
\newcommand{\specBS}{{\tt idlspec2d}}

\newcommand{\Sersic}{S\'ersic}
\begin{document}
\title{The Eighth Data Release of the Sloan Digital Sky Survey: First
  Data from SDSS-III}

\author{
Hiroaki Aihara\altaffilmark{1},
Carlos Allende Prieto\altaffilmark{2,3},
Deokkeun An\altaffilmark{4},
Scott F. Anderson\altaffilmark{5},
\'Eric Aubourg\altaffilmark{6,7},
Eduardo Balbinot\altaffilmark{8,9}
Timothy C. Beers\altaffilmark{10},
Andreas A. Berlind\altaffilmark{11},
Steven J. Bickerton\altaffilmark{12}, 
Dmitry Bizyaev\altaffilmark{13}
% ...
}

\altaffiltext{1}{
Institute for the Physics and Mathematics of the Universe,
The University of Tokyo,
5-1-5 Kashiwanoha, Kashiwa, 277-8583, Japan.
}

\altaffiltext{2}{
Instituto de Astrof\'\i{}sica de Canarias, E38205 La Laguna, Tenerife, Spain.
}

\altaffiltext{3}{
Departamento de Astrof\'{\i}sica, Universidad de La Laguna, 38206, La 
Laguna, Tenerife, Spain.
}

\altaffiltext{4}{
Department of Science Education,
Ewha Womans University, Seoul 120-750, Korea.
}

\altaffiltext{5}{
Department of Astronomy, University of Washington, Box 351580, Seattle, WA
98195.
}

\altaffiltext{6}{
Astroparticule et Cosmologie (APC), Universit\'e Paris-Diderot, 10 rue
Alice Domon et L\'eonie Duquet, 75205 Paris Cedex 13, France.
}

\altaffiltext{7}{
CEA, Centre de Saclay, Irfu/SPP,  F-91191 Gif-sur-Yvette, France.
}

\altaffiltext{8}{
Instituto de F\'\i sica, UFRGS, Caixa Postal 15051, Porto Alegre, RS -  
91501-970, Brazil.
}

\altaffiltext{9}{
Laborat\'orio Interinstitucional de e-Astronomia, - LIneA, Rua
Gal.~Jos\'e Cristino 77, Rio de Janeiro, RJ - 20921-400, Brazil.  
}

\altaffiltext{10}{
Department of Physics \& Astronomy
and JINA: Joint Institute for Nuclear Astrophysics, Michigan
State University, E. Lansing, MI  48824.
}

\altaffiltext{11}{
Department of Physics and Astronomy, Vanderbilt University, Nashville  
TN 37235.}

\altaffiltext{12}{
Department of Astrophysical Sciences, Princeton University, Princeton, NJ
08544.
}

\altaffiltext{13}{
Apache Point Observatory, P.O. Box 59, Sunspot, NM 88349.
}

% ...

\shorttitle{SDSS DR8}

\begin{abstract}
The Sloan Digital Sky Survey (SDSS) started a new phase in August
2008, with new instrumentation and new surveys focused on Galactic
structure and chemical evolution, measurements of the baryon
oscillation feature in the clustering of galaxies and the quasar
Ly$\alpha$ forest, and a radial velocity search for planets around
$\sim 8000$ stars.  This paper describes the first data release of
SDSS-III (and the eighth counting from the beginning of the SDSS).
The release includes five-band imaging of roughly 5200 deg$^2$ in the
Southern Galactic Cap, bringing the total footprint of the SDSS
imaging to 14,555 deg$^2$, or over a third of the Celestial Sphere.
All the imaging data have been reprocessed with an improved
sky-subtraction algorithm and a final, self-consistent photometric recalibration
and flat-field determination.  This release also includes all data
from the second phase of the Sloan Extension for Galactic
Understanding and Exploration (SEGUE-2), consisting of spectroscopy of
approximately 118,000 stars at both high and low Galactic
latitudes. All the more than half a million stellar spectra obtained
with the SDSS spectrograph have been reprocessed through an improved
stellar parameters pipeline, which has better determination of
metallicity for high metallicity stars.
\end{abstract}

\section{Introduction}
The Sloan Digital Sky Survey (SDSS; York \etal\ 2000) saw first light
in May 1998, and has been in routine survey operation mode since April
2000.  It uses a 2.5m telescope with an unvignetted $3^\circ$
field of view (Gunn \etal\ 2006) at Apache Point Observatory (APO) in
Southern New Mexico, which is dedicated to wide-angle
surveys of the sky.  The first and second phases of the survey (SDSS-I
and SDSS-II) were carried out with two instruments: a drift-scan
imaging camera (Gunn \etal\ 1998) with 30 CCDs imaging in five filters
($ugriz$, Fukugita \etal\ 1996), and a pair of double spectrographs, fed by 640
optical fibers.  The imaging data, essentially all of which have been taken
under photometric and good-seeing conditions (Ivezi\'c \etal\ 2004;
Padmanabhan \etal\ 2008; see also Hogg \etal\ 2001), now cover
more than 14,500 deg$^2$ in five filters (of which about 11,600
deg$^2$ was observed as part of SDSS-I/II), or roughly one third of
the Celestial Sphere.  The 50\% completeness limit for point sources is
$r=22.5$.  The data have been analyzed with a
sophisticated pipeline (Lupton \etal\ 2001) and have been
photometrically (Tucker \etal\ 2006, Padmanabhan \etal\ 2008; see also
Smith \etal\ 2002) and astrometrically (Pier \etal\ 2003) calibrated;
the resulting catalog contains almost half a billion distinct detected 
objects.  Well-defined samples of galaxies (Strauss \etal\ 2002;
Eisenstein \etal\ 2001), quasars (Richards \etal\ 2002a), stars (Yanny
\etal\ 2009) and other
objects are selected for spectroscopy; the survey has
obtained roughly 1.8 million spectra of galaxies, stars, and quasars
as of Summer 2009.  

The principal scientific goal of SDSS-I (2000--2005) and much of
SDSS-II (2005--2008) was to create a well-calibrated and contiguous
imaging and spectroscopic survey of the Northern Galactic Cap at high
Galactic latitudes, with the spectroscopy primarily focused on
extragalactic targets. We refer to this project in what follows as the Legacy
Survey.  SDSS-II carried out two additional surveys.  The Sloan
Extension for Galactic Understanding and Exploration (SEGUE; Yanny
\etal\ 2009) imaged a series of stripes sampling low
Galactic latitudes (each 2.5$^\circ$ wide and tens to hundreds of
degrees long), together with spectroscopy of roughly 250,000 stars, to
study Galactic structure, dynamics, and chemical composition.  The
SDSS Supernova Survey (Frieman \etal\ 2008) used approximately 80 repeat
scans of a $2.5^\circ \times 100^\circ$ stripe centered on the
Celestial Equator in the Southern Galactic Cap to identify Type Ia
supernovae with redshifts less than about 0.4, and to use them as
cosmological probes (Kessler \etal\ 2009); almost 500 objects
were spectroscopically confirmed as Type Ia supernovae. 

These data have been made public in a series of yearly data releases
(Stoughton \etal\ 2002; Abazajian \etal\ 2003, 2004, 2005, 2006;
Adelman-McCarthy \etal\ 2007, 2008; Abazajian \etal\ 2009; hereafter
the EDR, DR1, DR2, DR3, DR4, DR5, DR6, and DR7 papers, respectively).
These data have been used in over 3500 refereed papers to date for
studies ranging from asteroids in the Solar System to the discovery of
the most distant quasars.

It was clear, as SDSS-II was nearing completion, that the wide-field
spectroscopic capability of the SDSS telescope and system remained
state-of-the-art, and a new collaboration was established to carry
out further surveys with this telescope.  This new phase, called
SDSS-III, consists of four interlocking
surveys; it is described in detail in a companion paper (Eisenstein
\etal\ 2011).  In brief, these surveys are: 
\begin{itemize} 
\item SEGUE-2.  This survey is an extension of the spectroscopic component
  of the SEGUE survey of SDSS-II, extending the survey footprint in area and
  using revised target selection to increase the number of spectra in
  the distant halo of the Milky Way. 
   SEGUE-2 used the SDSS-I/II
  spectrograph and ran from August 2008 through July 2009. 
\item The Baryon Oscillation Spectroscopic Survey (BOSS).  This survey will
  measure the baryon oscillation signature in the correlation function
  of galaxies and the quasar Lyman $\alpha$ forest.  BOSS started operations
  in Fall 2009, and consists of a
  redshift survey over 10,000 deg$^2$ of 1.5 million luminous red galaxies to $z \sim
  0.7$, together with spectroscopy of 150,000 quasars with $z > 2.2$.
  This has required increasing the imaging footprint of the survey,
  and we have obtained an additional $\sim 2500$ deg$^2$ 
  of imaging data in the Southern Galactic Cap using the SDSS imaging
  camera.  In addition, in Summer 2009 the SDSS spectrographs
  underwent a major upgrade (new gratings, new CCDs, and new fibers) to
  improve their throughput and to increase the number of fibers from
  640 to 1000.
\item The Multi-object APO Radial Velocity Exoplanet Large-area Survey
  (MARVELS) uses a fiber-fed interferometric spectrograph that can
  observe sixty objects simultaneously to obtain radial velocities 
  accurate to 10--40 m s$^{-1}$ for stars with $9 < V< 12$.  Each star will be observed roughly 24
  times in a search for extrasolar planets.  The instrument has been
  in operation since Fall 2008.
\item The Apache Point Observatory Galactic Evolution Experiment
  (APOGEE) will use a fiber-fed H-band spectrograph with a resolution of 30,000,
  capable of observing 300 objects at a time.  The spectrograph will see
  first light in 2011, and will obtain high signal-to-noise
  ratio (S/N) spectra of roughly 100,000 stars in a variety of
  Galactic environments, selected from the Two-Micron
  All-Sky Survey (2MASS; Skrutskie \etal\ 2006). 
\end{itemize}

SDSS-III started operations in August 2008 and will continue through
July 2014.  As with SDSS-I/II, the data will periodically be released
publicly; this paper describes the first of these releases.  For continuity with the
previous data releases of SDSS-I/II, we refer
to it as the eighth data release, DR8.  DR8 includes two significant items
of new data relative to DR7:
\begin{itemize} 
\item Roughly 2500 deg$^2$ of imaging data in the Southern Galactic Cap,
  taken as part of BOSS. 
\item SEGUE-2 spectroscopy, consisting of 204 unique plates with
  spectra of roughly 
  118,000 stars. 
\end{itemize}

As with previous data releases, DR8 is cumulative, and includes
essentially all
data from the previous releases.  However, this is not just a repeat of
previous data releases, but also an enhancement.  In particular, we
have re-processed all SDSS-I/II imaging data using a new version of
the imaging pipeline with a more sophisticated sky subtraction
algorithm, and all stellar spectra have been re-processed with an
improved stellar parameters pipeline.  

This paper provides an overview of DR8.  Section~\ref{sec:scope}
describes the scope of the imaging and spectroscopic data. More
details on the changes to the photometric pipeline and photometric
calibration may be found in \S\ref{sec:photo}, while the
spectroscopy, including SEGUE-2 target selection, is described in
\S\ref{sec:spectro}.  Methods for accessing these data are presented
in \S\ref{sec:data}.  We conclude, and outline the plan for future
SDSS-III data releases, in \S\ref{sec:summary}.  The data, and
portals to access them, are described in greater detail at the DR8
website\footnote{\tt http://www.sdss3.org/dr8/}. 

\section{Scope of DR8}
\label{sec:scope} 

\begin{figure}[t]\centering\includegraphics[width=10cm]{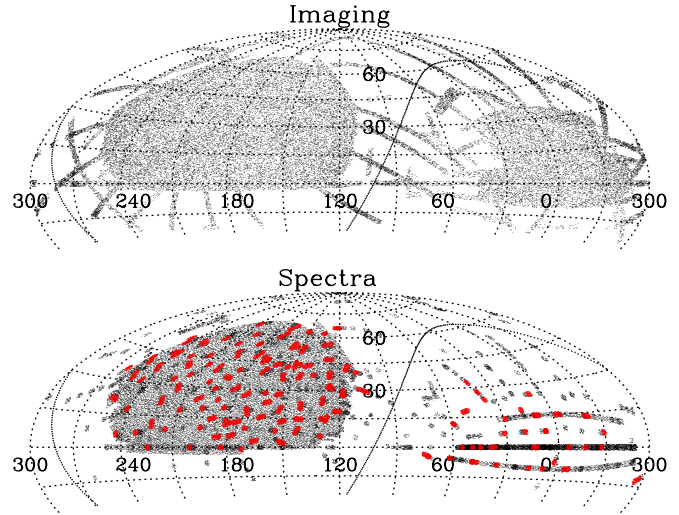}
\caption{
The sky coverage of DR8 in J2000 Equatorial coordinates, in imaging (upper)
and spectroscopy (lower).  Right ascension $\alpha = 120^\circ$ is at the
center of these plots.  The Galactic plane is the solid curve that
snakes through the figure. Note the
contiguous imaging coverage of the Southern Galactic Cap (centered
roughly at $\alpha = 0^\circ, \delta = +10^\circ$); in DR7, this
region of sky was covered by a few disjoint stripes.  The red regions
in the lower panel show the coverage of the SEGUE-2 plates.  The BOSS
survey will obtain spectra over 10,000 deg$^2$, including the
contiguous areas in the Northern and Southern Galactic caps. 
\label{fig:sky_coverage}}
\end{figure}

The contents and sky coverage of the data release are summarized in Table~\ref{table:dr8_contents} and
Figure~\ref{fig:sky_coverage}.  The principal change in the imaging
footprint from that in DR7 is the coverage of a large contiguous region, 3172 deg$^2$,
in the Southern Galactic Cap.  Three disjoint stripes (76, 82,
and 86, centered roughly at $\alpha = 0^h, \delta = -10^\circ,
0^\circ$, and $+15^\circ$, respectively) were included in DR7.  The
remaining area, roughly 2500 deg$^2$, was
observed in the Fall and early Winter months of 2008 and 2009; it will
be used to identify spectroscopic targets for the BOSS survey.
Including the SEGUE stripes, the total area in the Southern Galactic
Cap is 5194 deg$^2$.  

The total sky coverage of DR8 has been
calculated more carefully than was done with DR7, so the
solid angle coverage of the two cannot be quite directly compared. 
The 
figure and the sky coverage numbers do not distinguish some of
the ``special'' scans described in previous data release papers.
In particular, the scans covering M31 (DR5 paper), the Orion
region (Finkbeiner \etal\ 2004), and the SEGUE-1 imaging scans (Yanny
\etal\ 2009) are all represented in the figure, and are included
in the data release along with the Legacy imaging in the same files
and database tables.

On the spectroscopic side, the footprint of the survey has increased
only slightly, given the small number of SEGUE-2 plates that lie
outside the contiguous area of the North Galactic Cap (see the red
regions in Figure~\ref{fig:sky_coverage})\footnote{The solid
  angle listed in the DR7 paper for the spectroscopic footprint
  added together the Legacy and SEGUE-1 areas, double-counting 
  the overlap between the two.}.  The numbers of spectra included in
various classifications are based on \specBS\ (occasionally referred
to as ``specBS''; see \S\ref{sec:specBS}), one of the two pipelines
used in DR7 to classify spectra and determine redshifts.  Note that
unlike Table 1 in the DR7 paper, this table lists only those {\em unique}
spectra (i.e., duplicates have been removed), for which \specBS\ gave
no redshift warning flags other than {\tt MANY\_OUTLIERS} (see Table 4
of the DR6 paper).  Furthermore, the DR7 paper based its numbers on the results
of the other of these pipelines, {\tt spectro1d} (Subbarao \etal\ 2002), 
but comparisons of the two pipelines (DR6 paper) show that they
are in substantive agreement for over 98\% of spectra. 

The {\tt idlspec2d} classifications are assigned automatically and do
not include the results of any eyeball inspection.  This fact, and the
absence of a luminosity cut in the definition of quasars, means that
the number of quasars differs somewhat from the DR7 Quasar Catalog
(Schneider \etal\ 2010).  Objects listed as ``unclassifiable'' in
Table~\ref{table:dr8_contents} are sources with spectroscopic
classification warning flags: most such objects have low
signal-to-noise ratio or problems with the data (e.g., due to bad
columns), but this category also includes unusual objects with extreme
properties, such as featureless BL Lacertae objects (e.g., Collinge
\etal\ 2005; Plotkin \etal\ 2010), extreme broad absorption-line
quasars (e.g., Hall \etal\ 2002) or unusual types of metal-rich or magnetic
white dwarfs (e.g., Dufour \etal\ 2007; Schmidt \etal\ 2003).

\begin{deluxetable}{lr}
\tablecaption{Coverage and Contents of DR8
              \label{table:dr8_contents}}

\startdata

\cutinhead{\bf Imaging} 

 Total unique imaging area covered &14,555 deg$^2$\\
 Total area imaged, including overlaps\tablenotemark{a} & 31,637 deg$^2$\\
 New imaging area since DR7 & $\sim 2500$ deg$^2$ \\
 Unique objects in database & 469,053,874\\
\cutinhead{\bf Spectroscopy}
 Spectroscopic footprint area\tablenotemark{b} &9274 deg$^2$\\
 \quad Legacy &7966 deg$^2$\\
 \quad SEGUE-1  &1424 deg$^2$\\
 \quad SEGUE-2 & 1317 deg$^2$\\
 Total number of plate observations\tablenotemark{c} & 2880 \\
 \quad Legacy survey plates\tablenotemark{c} & 1926 \\
 \quad Special plates\tablenotemark{c} & 301 \\
 \quad SEGUE-1 survey plates\tablenotemark{c} & 442 \\
 \quad SEGUE-2 survey plates\tablenotemark{c} & 211 \\
 Total number of spectra\tablenotemark{d}& 1,629,129\\
 \quad  Galaxies   &  860,836 \\
 \quad  Quasars    & 116,003 \\
 \quad  Stars      &  521,990\\

 \quad  Sky        & 93,187 \\
 \quad  Unclassified\tablenotemark{e} & 37,113\\ 
\enddata
\tablenotetext{a}{Includes only some of the repeat scans on Stripe 82 taken
  in 2005-2007 as part of the SDSS supernova survey.  Roughly 50\% of
  the SDSS footprint has been imaged more than once.}
\tablenotetext{b}{This area does not double-count the overlapping
  footprint of the Legacy and SEGUE surveys.}
\tablenotetext{c}{Each plate has 640 fibers.  The number of plates
  includes some repeat observations.}
\tablenotetext{d}{Spectral classifications from the \specBS\ 
  code; the totals do {\it not} include duplicates or spectra with
  redshift warning flags. }
\tablenotetext{e}{I.e., objects in which {\tt ZWARNING} (DR6 paper, Table 4) have any bit
  other than {\tt MANY\_OUTLIERS} set.} 
\end{deluxetable}

\section{Imaging Data}
\label{sec:photo}

DR8 includes essentially all the DR7 
data, together with the additional data described above.
The major exceptions to this statement are as follows:
\begin{itemize} 
\item Some of the SEGUE-1 imaging scans described in the DR7 paper
  pass through the Galactic plane, where the SDSS photometric pipeline
  does a poor job in regions of very high stellar density.  We
  also processed these fields with software from the Pan-STARRS (Kaiser
  2002) collaboration.  The results of that analysis are still
  available on the DR7 website, and we do not separately make them
  available in DR8.  However, DR8 {\em does} include the SDSS photometric
  pipeline results in these regions; at sufficiently low latitudes,
  where the stellar density exceeds 5000 stars deg$^{-2}$ brighter
  than $r = 21$, the SDSS photometry is likely to be unreliable.  In particular, there are
  regions of sky that are so crowded that the software simply times
  out, and no objects are included in the catalog.   This effect is
  visible in Figure~\ref{fig:sky_coverage} as the discrete scans near
  $\alpha = 300^\circ$   that simply fade away at low latitudes.  
\item DR8 does not include the coaddition of the repeat scans on
  Stripe 82 (see the DR7 paper), and it includes only some of the
  Stripe 82 runs (often taken under 
  non-photometric conditions) obtained as part of the SDSS Supernova
  Survey.  In particular, in the resolving (\S\ref{sec:resolve}) of
  Stripe 82, we identified the highest quality run (via the ``score''
  value described in \S\ref{sec:score}) at each position.  We include
  the entire run in DR8 if it is the highest quality
  at at least one point in the stripe.  DR8 includes 118 runs in total
  on Stripe 82.   {\em All} 303 Stripe 82 runs are available in DR7, making
  DR7 the dataset to be used for analyses of time-variable phenomena
  in the stripe.  
\item The DR4 paper (see also Ivezi\'c \etal\ 2004) describes web pages documenting detailed
  diagnostics of the photometric and astrometric quality on a run-by-run
  basis, based both on internal consistency checks and overlaps between
  adjacent runs.  These remain on the DR7 website;
  we have not repeated this analysis for the reprocessing of the imaging
  data for DR8  or for the new data from Fall 2008 or later.
  Note that the
  ``ubercalibration'' procedure described in \S\ref{sec:ubercal} does
  explicitly report the reproducibility of photometry in overlapping
  runs.  The documentation on the DR8 web site describes how to check those
  results.
\end{itemize}

In the following subsections, we outline further differences
to the image processing relative to DR7, including updates to
the sky subtraction algorithm (\S\ref{sec:sky} and \S\ref{sec:globalsky}), photometric
calibration (\S\ref{sec:ubercal}), resolving overlapping runs
(\S\ref{sec:resolve}), and astrometric calibration
(\S\ref{sec:astrom}).   We also describe the availability of galaxy
morphologies from the Galaxy Zoo collaboration
(\S\ref{sec:galaxyzoo}). 

\subsection{Improved Sky Subtraction}
\label{sec:sky}

The SDSS imaging data are all processed with the Photometric Pipeline
(\photo).  A number of investigators have shown that the sky subtraction
algorithm used by the DR7 photometric pipeline causes it to systematically
underestimate the brightness of large galaxies (Blanton \etal\ 2005,
Lisker \etal\ 2006, Lauer \etal\ 2007, Bernardi \etal\ 2007, West
\etal\ 2010, 
among others; see also the discussion in the DR4, DR6, and DR7
papers).  The sense of the error was to oversubtract the outer
regions of large galaxies in the sky estimation, affecting the photometry both of those
galaxies and that of smaller and fainter objects in their vicinity.  The
DR8 imaging data were processed with a more sophisticated sky
subtraction algorithm that reduces this problem, but by no means
solves it completely. 

{\tt Photo} estimates the sky level on a rectangular grid of 128 pixels (roughly
$50''$) by calculating the median of the $256 \times 256$ pixels
centered on each grid point.  The version of \photo\ used in DR7 and
earlier data releases simply interpolated bilinearly between these grid points as an
estimate of sky; this approach tended to erroneously include light from
extended regions around bright galaxies, and thus underestimated
their fluxes.

The new algorithm adds an additional step of identifying and modeling these extended galaxies
before estimating the final sky level.  As described in Lupton
\etal\ (2001) and the EDR paper, \photo\ first estimates a single
preliminary sky value for an entire $10' \times 13'$
field\footnote{For a definition and explanation of the SDSS fields,
  see the EDR paper.}.  Using
this sky value, it identifies {\tt BRIGHT} sources ($> 51 \sigma$,
corresponding roughly to a star with $r = 20$).
These sources are next run through the deblender, to separate
overlapping {\tt BRIGHT} objects.  This step is new to this version of
\photo\ (before, the deblender was only run after the final sky model
was determined).   Models are determined
for each child object, and these are then subtracted from the frame.  The EDR paper
describes the models that are used for galaxies: two-dimensional
exponential and de Vaucouleurs profiles of arbitrary axis ratio,
convolved with the local PSF.  As described in the DR2 paper, one can
fit the observed profile of galaxies with a {\em linear combination}
of the best-fit exponential and de Vaucouleurs models to any given
galaxy in a given band; we refer to this as the ``cmodel''.   This
model is then subtracted from the image, removing the extended wings
of the galaxy. 

Unsaturated {\tt BRIGHT} stars are not subtracted at this stage.
However, for saturated stars, the outer wings 
(i.e., outside a radius of $28.2''$) are fit to a power law, of
index $\beta = -3.25$ in $ugrz$ and $\beta = -2.5$ in $i$\footnote{A
  small fraction of the photons scatter within the thick chips used in
  the $i$ band, yielding an extended halo around stars.}; these wings 
are then subtracted from the image.
Now that the wings of bright galaxies and saturated stars have been
subtracted, the local sky is estimated as before; that is, a clipped
median is measured on a 128-pixel grid and linearly interpolated. 

The galaxies (but {\em not} the stars\footnote{Not adding the stars
  back in greatly simplifies the deblending around bright stars, which
  otherwise cause significant parts of the frame to blend into a
  single object.}) are
then added back to the sky-subtracted frame, and faint object
detection proceeds, as described in the EDR paper.  Flags are set to
the mask image indicating that a significant part of the sky
background at that pixel came from nearby bright objects.  If
{\tt SUBTRACTED} is set (flux subtracted is more than $1\,\sigma$ above
the sky) the pixel is probably trustworthy, while {\tt NOTCHECKED}
pixels (more than $5\,\sigma$ above the sky) are probably unreliable
(and no further objects will be detected in these regions; the {\tt BRIGHT}
objects will of course be preserved).

With this change in the sky subtraction routine, the outer parts of
galaxies are considerably more extended than they were in the previous
version of the software, meaning that they are likely to overlap with
more objects in their outer parts.  With this in mind, we increase the
number of children any blended parent can be decomposed into from 25
to 100.  This has the negative effect of increasing the processing
time for fields in which there is a great deal of overlap between
objects, such as those at low latitudes and those with bright stars.
We find that \photo\ times out on 0.5\% of the fields at $|b| >
15^\circ$ (45 deg$^2$ in all), almost all of which have a particularly
bright star in the field.

\subsubsection{Photometry of Bright Galaxies}

\begin{figure}[t]\centering\includegraphics[width=10cm]
{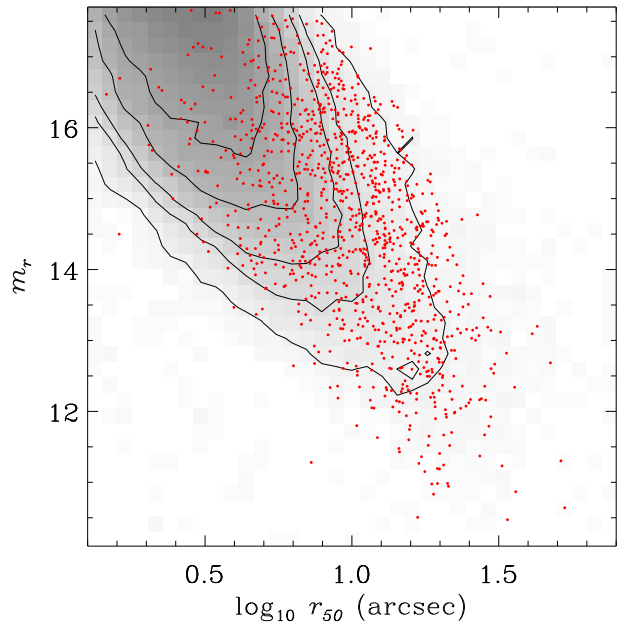}  
\caption{The grayscale and contours show the distribution of galaxies
  in SDSS in apparent magnitude and Petrosian half-light radius.  The
  red dots show the distribution of artificial galaxies added to the
  imaging frames to explore the ability of the pipeline to photometer
  large galaxies.  We have deliberately biased the sample of
  artificial galaxies to larger
  objects at a given magnitude. 
\label{fig:fakedist}}\end{figure}

We quantified the accuracy of bright galaxy photometry by adding 1300 artificial galaxies at
random positions to SDSS imaging frames, processing them with both the
old (DR7) and new (DR8) versions of \photo, and comparing the results
with the true input values.  The simulated galaxies, which have \Sersic\ radial profiles with
a range of inclinations and \Sersic\ indices, follow the observed
correlation between apparent magnitude and angular size seen for real
galaxies (Figure~\ref{fig:fakedist}). However, we biased the sample somewhat to larger and
brighter objects, as this is the regime in which the sky subtraction
errors are likely to be worst.  In addition, the sample is approximately
size-limited at a Petrosian (1976) half-light radius $r_{50} \sim 5''$.  

\begin{figure}[t]\centering\includegraphics[width=10cm]{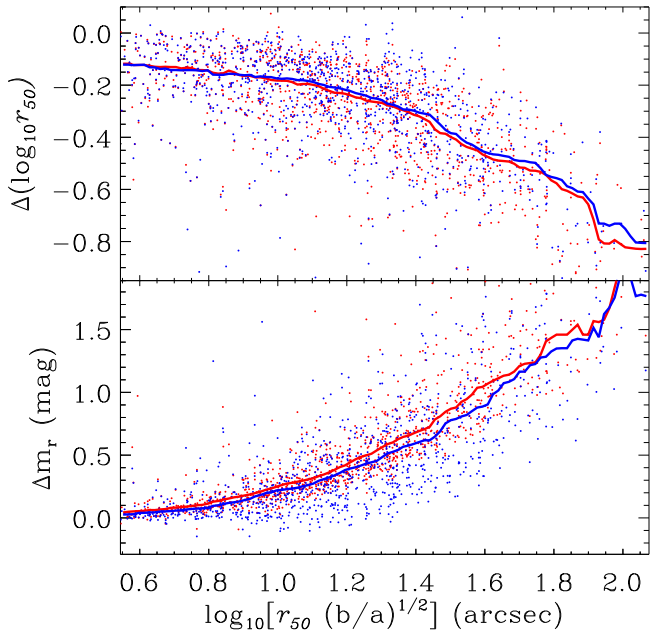}
\caption{ \label{fig:offsets} Differences between the true and measured  
  $r$-band half-light radii and magnitudes as a function of
  $r_{50}\times (b/a)^{1/2}$ (whose square is proportional to the area
  of the galaxy; here $b/a$ is the axis ratio of the galaxy from the
  model fit), for a sample of simulated
  galaxies.  The sample has \Sersic\ profiles, with a range of
  magnitudes and sizes (and therefore surface brightnesses), designed
  to sample the observed distribution of large bright galaxies.  The
  measured magnitudes are the combined ``cmodel'' magnitudes using the
  exponential and de Vaucouleurs fits, and the measured sizes are the
  effective radii from the better of those two fits for each galaxy. Top
  panel shows the logarithmic difference between the measured
  half-light radius and the true one ($\Delta \log_{10} r_{50} =
  \log_{10} r_{50,meas} - \log_{10} r_{50,true}$).  Bottom panel shows
  the magnitude difference ($\Delta m = m_{meas}-m_{true}$). Results
  are shown both for the version of \photo\ used in DR7 (red) and DR8
  (blue).  The running median values as function of radius are shown as the solid lines.  The new code
  reduces the bias at large area, but only incrementally.
\label{fig:dr8_offsets}}\end{figure}

The results are shown in Figure~\ref{fig:dr8_offsets}, where we plot
the difference between measured and true half-light radii and magnitudes in the $r$ band for
the simulated galaxies in the DR7 (red) and DR8 (blue) versions of
the pipeline.  Results for the other bands are similar.  The new sky
subtraction algorithm improves things somewhat, but is not a panacea.  The principal
trend is with galaxy area, because it is the quantity that couples most directly to the
sky measurement.
 The
improvement is subtle at best,
and is only visible for galaxies with $r_{50} > 30''$.  The roughly 1
mag of bias at $r_{50} \sim 50''$ is reduced in the DR8 pipeline by only
about 0.25 mag.  Additionally,
there is a distinct bias in the measured sizes themselves, which is
similar in the two pipelines. Some of the problem may not be due 
to sky subtraction, but rather to the deblender systematically
assigning some of the light in the outer parts of galaxies to
superposed fainter stars and galaxies.

\subsubsection{Photometry of Faint Galaxies near Bright Galaxies}

\begin{figure}[t]\centering\includegraphics[width=9cm]{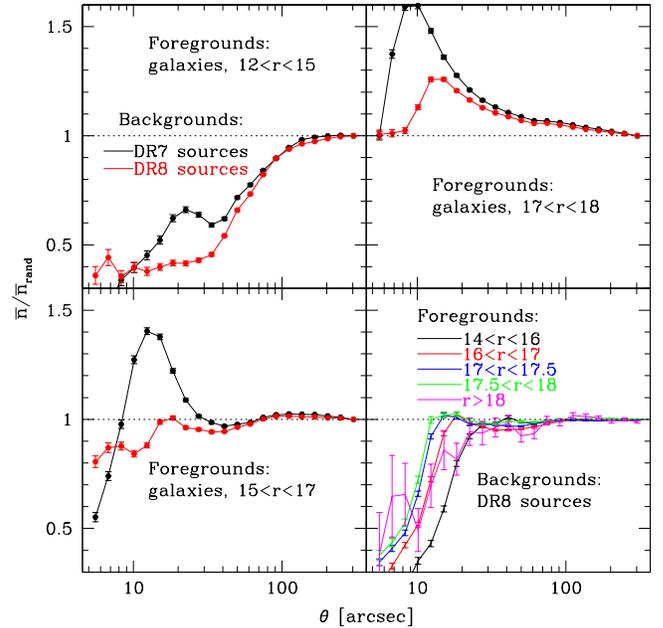}\caption{
{\em Top left, bottom left, top right:} Number density of source
galaxies as a function of distance from bright foreground galaxies.  Each panel is a separate
foreground magnitude bin as labeled on the plot.  The black solid and
red dashed lines show the results for DR7 and DR8, respectively.  {\em
  Bottom right:} Same as other panels but for DR8 only, where separate
line colors and styles indicate different foreground magnitude bins.
In this case, unlike for the other panels, source galaxy photometric
redshifts were used to exclude sources that are in front of, or are
physically associated with, the foreground object.
\label{fig:rachel_phot}}\end{figure}

A related problem reported by Mandelbaum \etal\ (2005) is that the previous
sky-subtraction procedure suppressed the number density
of faint galaxies around bright galaxies and distorted
the measured shapes of these faint galaxies, which affects 
measurements of galaxy-galaxy lensing and the clustering of faint
objects near bright objects.  We here examine the
suppression in the number density, comparing the DR7 and DR8
pipelines.  Figure~\ref{fig:rachel_phot} compares the number density
of faint galaxies 
relative to the mean in the two versions of the pipeline, as a
function of the angular distance from bright galaxies.

The upper panels and lower left panel of Figure~\ref{fig:rachel_phot}
show a test that used a common set of foreground galaxies
($12<r_{model}<18$), divided into magnitude bins.  The faint galaxies
came from the original catalog of source galaxies in Mandelbaum \etal\
(2005), which included well-resolved galaxies with $r<21.8$ selected
from the DR7 reductions.  For DR8, we selected a similar catalog of
source galaxies described in Reyes \etal\ (2011).  In these panels, we
did not attempt to exclude source galaxies that are physically
associated with the lens, which means that we expect some increase in
the number density at small scales where galaxy clustering is
important.  Additional effects that should modify the number density
include deblending errors around the bright galaxy, gravitational
magnification\footnote{Based on the lensing shear that is measured and
  the slope of the number counts of the source sample, we anticipate
  an effect that is at most 3\% at $10''$, is strictly positive, and
  decreases with scale.} and dust extinction (which tends to
counteract magnification but appears to be weaker for low redshift
galaxies; M\'enard \etal\ 2010).

As shown in the three aforementioned panels of
Figure~\ref{fig:rachel_phot}, the number density of faint galaxies 
around bright foreground galaxies is strongly affected by the foreground
at angular separations less than 100$''$.  The $12<r<15$ galaxies have such a large angular
extent that the number density is severely suppressed below 50$''$.
The sky mis-estimation near $15<r<17$ galaxies causes a $\sim 5$\%
suppression in the number density for $30<\theta<90''$.  Finally, for
the $17<r<18$ foreground galaxies, the predominant effect in the source number
density is clustering, but there is a subtle effect around 50$''$ that
is likely due to sky mis-estimation.  In all three panels, the curves
for the previous reductions exhibit a significant bump around
$20''$, the origin of which is unclear.  This bump is present around
stars as well (for which lens-source clustering is not a possible
explanation), but in the DR8 reductions, the bump goes away almost
completely for both stars and galaxies. The disappearance of this
artifact at $20''$ constitutes a substantial improvement in the new
pipeline.  Unfortunately, the suppression in source counts from
$40''<\theta<90''$ has improved only slightly.

The lower right panel in Figure~\ref{fig:rachel_phot} shows the results of a different test, using the new  
source catalog from the DR8 reductions only.  For this catalog, we  
have generated photometric redshifts for all sources using ZEBRA  
(Feldmann \etal\ 2006; Nakajima \etal\ 2011);  
these photometric redshifts were used to isolate sources at $z>0.3$, thus  
eliminating almost completely the correlations between foregrounds and  
sources due to galaxy clustering.  The remaining effects in the number  
density are due to sky subtraction, gravitational magnification, dust extinction, and possibly  
a very low level ($<2$\%) of clustering due to catastrophic photo-$z$  
errors.  As shown here, the sky subtraction suppresses the source
number density by $\sim 4\%$ for $30''<\theta<90''$.
Note that extended dust halos around galaxies (M\'enard \etal\ 2010)
cannot be the explanation of the effect, as the suppression is seen around stars
(not shown) as well as galaxies.  

The magnitude of the galaxy number suppression depends not just on the properties of the  
bright foreground galaxy (as illustrated in this figure), but also on  
the properties of the fainter nearby galaxies, with fainter or lower  
surface brightness galaxies being more severely affected.  Position on  
the CCD is also a factor: near the edges of the fields, the sky level  
must be extrapolated, which means that sky estimates are worse within
256 pixels of the edge.

\subsection{Improved Sky Subtraction in Post-Processing}
\label{sec:globalsky}

DR8 includes ``corrected frames'', FITS files of each frame which
have been bias subtracted and flat-fielded, with bad columns and
cosmic rays interpolated over.  Each frame has a World Coordinate
System (WCS) giving the full astrometric solution in its
header, and the pixel values are calibrated to fluxes.  Thus,
astrometry and photometry can be performed directly on the image.
These images have also been sky-subtracted, using an algorithm that
goes beyond the one we have described above. But the photometric
pipeline has {\em not} been run on these corrected frames, as we
implemented this fix after the processing of the bulk of the data was
completed, thus
these improvements are not reflected in the object catalogs.

Our method treats each run as a whole, and fits a smooth function to
the variation of the sky background using a heavily masked and binned
image of the run.  The method is described in full by Blanton \etal\
(2011).  We find good agreement within the typical image noise between
the photometry of point sources in these images and the SDSS catalog.

More critically, we have also tested the effect of our background
subtraction on the photometry of large galaxies by inserting fake
galaxies into the raw data and measuring their properties after
background subtraction.  We find that this sky-subtraction technique
introduces biases of $>0.1$ mag only at half-light radii
$r_{50}>100''$.  For typical large galaxies, our results agree at the
5\% level with those of the Montage package distributed by the
NASA/IPAC Infrared Science Archive, which uses overlapping
observations from adjacent runs to determine the sky levels (Berriman
\etal\ 2003).  However, any actual photometry of such galaxies is much
more difficult, requiring very accurate deblending as well to achieve
unbiased results. These issues are more fully explored in Blanton
\etal\ (2011).  See the paper by West \etal\ (2010) for another
approach to the problem.  

We recommend these sky-subtracted images as a robust starting point
for users interested in reprocessing SDSS images.  Note that
for very large systems (for example for intracluster light studies)
there may still be biases present. For this reason, the corrected
frames also contain the information needed to undo the 
global sky subtraction.

\subsection{Photometric Calibration}
\label{sec:ubercal}
 
In SDSS-I/II, the default photometric
calibration method used an auxiliary 24-inch telescope (the ``Photometric
Telescope'', hereafter PT), which observed a set of standard stars
(Smith \etal\ 2002) to determine the photometricity and extinction
coefficients for each night, as well as a large set of calibration
fields on the stripes observed by the 2.5m to place them on a uniform
photometric system (Tucker \etal\ 2006).  While this approach allowed
us to reach our goal of 2\% rms photometric calibration in all bands
(Ivezi\'c \etal\ 2004), it was limited by concerns about the slightly
different photometric systems of the PT, the 2.5m, and the Naval
Observatory 1.0m telescope in Flagstaff, where the standard stars were
initially put onto a common system.  In addition, this approach did
not take advantage of the overlap between adjacent scans.  

An alternative approach, called ``ubercalibration'' (Padmanabhan
\etal\ 2008), is a purely internal calibration using only the overlaps between
adjacent scans of the 2.5m.  This new calibration is forced to be
on the same zeropoint (within 1 millimag in $griz$ and 3 millimag in
$u$) on average as the DR7 
calibration, but it does not use any data from the PT.  As described in
Padmanabhan \etal\ (2008), the calibration has residual errors of
order 1\% in $griz$ and 2\% in $u$.

Ubercalibration uses a series of scans running perpendicular to the
main survey runs, performed in a fast, binned mode available on the
SDSS camera, referred to as the Apache Wheel scans.  The uncalibrated
version of these data and their associated reductions are available as
flat files on the DR8 website, but their proper use requires a great
deal of care.

We made both PT calibration and ubercalibration results available in
DR6 and DR7, but with DR8, we release only the results based on the
ubercalibration.  In particular, the PT calibration was not performed
for the new imaging data.  The DR7 ubercalibration process used a
different flat-field scheme from that used in DR8; this difference
dominates the difference in the calibration between the two.

Schlafly \etal\ (2010) have used DR8 photometry to study the effects
of Galactic reddening on star colors (in particular, the blue tip of
the stellar locus); they find rms spatial variations in these colors
of 18, 12, 7, and 8 millimag in $u-g$, $g-r$, $r-i$, and $i-z$,
respectively.  These variations include possible contributions from
stellar population variations and errors in the Schlegel, Finkbeiner,
\& Davis (1998, SFD) dust map as well as photometric calibration
errors, and so represent {\em upper limits} on the amplitude of
the latter. Of course, these values are consistent with the 1\% rms calibration
errors quoted above in $g,r,i$ and $z$.  Schlafly \etal\ (2010) also find
systematic differences in zeropoints between the North and South
Galactic Cap, of 8, 22, 7, and 12 millimag in $u-g$, $g-r$, $r-i$, and
$i-z$, respectively (as Figure~\ref{fig:sky_coverage} shows, the North
and South are tied together photometrically with a few SEGUE imaging
scans).  Again, these differences may be due in part to errors in the
SFD map and stellar population differences.

  With the changes in \photo\ and calibration, it is interesting to
  compare the DR7 and DR8 photometry.  For a sample of $18 < r < 19.5$
  stars randomly selected over the DR7 footprint, we found the PSF
  magnitudes to differ by an rms of 11-14 millimag in $griz$.  In $u$
  band, we further restricted ourselves to $u < 20$, and found rms
  differences of 20 millimag. 

\subsection{Resolving the Imaging}
\label{sec:resolve}
The SDSS imaging camera (Gunn \etal\ 1998) observes the sky in six
parallel {\em scanlines}, each $13'$ wide and as much as hundreds of
degrees long.  As is discussed in detail in the EDR paper, the way the
camera scans the sky produces quite a bit of overlap between the scanlines. The geometry of the great circles of
the main SDSS survey naturally gives rise to substantial overlap at the
ends of the stripes (York \etal\ 2000); it is this overlap which allows the
photometry of the scans to be tied together (\S\ref{sec:ubercal}).
The overlap also allows accurate photometry of objects which may be close to
a CCD edge in one imaging run but far enough away to allow proper
measurement in the adjacent run.  
Roughly 50\% of the SDSS imaging footprint was observed more than
once, and the first two entries in Table~\ref{table:dr8_contents} show
that because of the overlaps, the total area imaged is more than double
the {\it unique} area.  

However, for statistical studies, one needs a single unique detection
of each object in the sky, which requires that we {\em resolve} the
overlaps, identifying a single imaging run to represent each point in
the SDSS imaging footprint.  In previous data releases, this was done
by simply bisecting the overlap between adjacent scanlines; the
primary detections of all objects lying on one side of the bisector
were assigned to one scanline, and those on the other side were
assigned to the other.  This procedure has several disadvantages that
motivated us to revisit the problem:
\begin{itemize} 
\item This approach makes most sense when the scans are all roughly parallel
  great circles, in the $\lambda,\eta$ coordinate system used by the
  Legacy survey (EDR paper; Pier \etal\ 2003).  It does not translate
  well to scans that use a different survey pole, such as the SEGUE imaging
  scans, the so-called oblique scans, and others.
\item Because of the focus on the Legacy survey in SDSS-I/II, the
  resolution of the scans made reference to the boundaries of an
  ellipse on the sky into which the Northern Galactic Cap scans
  approximately fit (York \etal\ 2000).  This meant that scans that
  happened to fall outside that ellipse were not flagged as
  ``primary'' (see below). 
\item Anticipating the possibility of further scanlines beyond the
  boundaries of the existing imaging runs, the resolve algorithm
  applied the bisector line (and thus flagged perfectly good
  imaging data as ``secondary''; see below) at the boundaries of the survey. 
\end{itemize}

In this section, we describe the new resolve algorithm.  We first
determine the geometrical sky coverage of the survey (or ``window
function,'' which describes which imaging data are primary at each
point of the survey footprint), then resolve it to produce a catalog
of primary, unique detections of objects.  The primary area of the
survey is constructed as a union of the individual SDSS fields, with the
highest scoring field covering any given point of the sky (in the
sense described in \S\ref{sec:score}) labeled as ``primary''.  We
will refer to detections in a non-primary part of a field as
``secondary'' if they are associated with a primary detection.  If
detections in non-primary areas are not associated with any primary
detection, we refer to them as ``best''.  Variable objects or those
close to the photometric limit of the catalog can give rise to such
unique detections in secondary observations of a given field.

\subsubsection{Scoring each field}
\label{sec:score}

As described in the EDR paper, the individual scanlines are divided 
into $10' \times 13'$ fields (1489 pixels by 2048 pixels), with 128
pixels of overlap between them.  The photometric pipeline analyzes
each field separately.  As a first step in defining the geometry of
the full survey, we trim 64 pixels (about $25''$) off each edge of the field.  This
removes the overlap between adjacent fields along the scanline, while
the trimming perpendicular to the drift scan direction prevents the
primary catalog from including objects that are too close to the frame
edge to be properly measured.

Each point on the sky can be covered by one or more fields, and we
need to identify the best of these to represent the field.  We do so
by first 
ranking the fields according to a metric which we refer
to as its ``score''.  This score is based on the
$r$-band seeing, the sky brightness in $r$, the measurement of
photometricity from ubercalibration and the APO $10\mu m$ cloud camera
(Hogg \etal\ 2001), and any indications of problems
when the imaging data were taken (poor focus or tracking, 
unusually high CCD noise or evidence that the flat-field petals were not
properly opened during the observations).  
Each field is given a numerical score between 0 and 1; values below 0.6 indicate that
the data are not photometric (as determined by the ubercalibration
process itself, and by the cloud camera).  These scores are used in
what follows to define the 
primary field covering each point on the sky.   

\subsubsection{Defining the window function}

The primary survey area is defined as the union of all the
fields.  Determining the window function requires identifying the
fields that cover each
position on the sky, and deciding which of those fields should
be considered primary at that position.

We treat each field as a rectangle on the sky defined by its trimmed
area as described above.  There is a unique set of disjoint polygons
(hereafter, ``balkans'') on the sky defined by all the field
boundaries, which are calculated using the {\tt mangle} package of
Swanson \etal\ (2008).  Each field is divided into one or more balkans,
and each balkan is fully covered by a unique combination of one or
more fields.  

We assign the primary field associated with each balkan as follows. We
start with the highest scored field overall, and call it primary for
all the balkans covered by it.  Then we step to its adjacent fields in
the same scanline.  As long as their score is within 0.05 of the
initial field, we consider them to be primary for the balkans they cover as
well; this avoids switching field-by-field between two comparably good
runs on the same scanline.  We continue along the scanline in both
directions until we reach a substantially worse field than the first
(i.e. a decrease in score of $>0.05$).  When that happens, we step to the
next highest ranked field that has not already been assigned, and
execute the same steps for that field.  Of course, if a balkan has
already been assigned a primary field, that assignment is not changed.  This
process is iterated until we have assigned all of the fields in the
survey.

\subsubsection{Resolving catalog detections}

Once the window function is defined, we can resolve multiple
detections of individual objects.  Each detection of an object 
has an associated flag, {\tt resolveStatus}, that reports the
results of this
procedure.  This exercise is performed only for those objects that are not
parents of deblended children, are not classified as {\tt BRIGHT}
detections (because they will be remeasured in a second pass through
the pipeline; see the EDR paper), and have not been classified as {\tt SKY}
(blank fields at which spectroscopic fibers can measure the spectrum
of the sky) or {\tt CR} (cosmic rays).  
We select objects which are in the full area of each field, excepting the 64
rows at the top and bottom (that is, those
overlapping adjacent fields on the same scanline).  Along those edges
in the drift scan direction, we take care to account for small
astrometric differences that might give rise to lost or duplicate objects: if
any two detections in adjacent fields are within $2''$ of each other
and straddle an edge, one and only one of them is chosen as primary
for the run.  The {\tt RUN\_PRIMARY} bit of {\tt resolveStatus} is set
for those objects that pass this cut.

We next define the ``survey primary'' detections, unique detections
among all the imaging runs.  In order to allow for small astrometric
jitter between adjacent balkans,
we select {\tt RUN\_PRIMARY} detections that are within the trimmed
area of the primary field covering the balkan, {\em or} within $1''$
of the edge of the balkan, and match each selected detection to the
current list of primary detections. If it matches a previous primary
detection, as it might if it is near the edge of the balkan, then it
is not included; otherwise, it is assigned {\tt SURVEY\_PRIMARY} in
{\tt resolveStatus}.

This process has the potential to miss some transient or low S/N
sources, which may be detected in some fields covering a region of sky
but not in the primary field.  To identify these,  
we loop
over all the fields, and match all the {\tt RUN\_PRIMARY} objects
to the full list of {\tt SURVEY\_PRIMARY} objects.  Objects that are
unmatched are good detections in this field, but have no corresponding
primary objects, and so fall into a separate category; we label them
as {\tt SURVEY\_BEST} in {\tt resolveStatus}.

Finally, the duplicate detections of primary or best objects are
called ``survey secondary'' detections.  To find these cases, we loop over
all fields and select objects which are {\tt RUN\_PRIMARY} but neither
{\tt SURVEY\_PRIMARY} nor {\tt SURVEY\_BEST}. We match these objects
against the {\tt SURVEY\_PRIMARY} and {\tt SURVEY\_BEST} lists from
the other fields.  If the detection is matched, and the balkan containing the
primary/best observation contains the current field we are
considering, then this detection is labeled {\tt SURVEY\_SECONDARY}.

This process produces a list of all of the primary,
best and secondary detections.  In addition, for each secondary
detection we know which primary or best detection matches it. The
documentation on the DR8 web site describes how to use this
information, which is useful for finding multiple
observations of the same object.

\subsection{Differences in Astrometric Calibration}
\label{sec:astrom}

The quality of the DR8 astrometry unfortunately is degraded from that in DR7 due to
a number of software errors introduced in the DR8
reprocessing.  The following effects apply to the DR8 astrometry:

\begin{itemize}

\item Color terms were not included in the transformation from
position on the detector to right ascension and declination.  This
causes 10-20 mas systematic errors with color in catalog
positions.  Systematic errors of similar size are introduced in 
the measure of position offsets between filters; 
the errors are somewhat smaller
in $i$ and $z$, and somewhat larger in $u$ and $g$.

\item The DR7 astrometry was calibrated against the Second US Naval
  Observatory CCD Astrograph Catalog (UCAC2; Zacharias \etal\ 2004).
  The UCAC2 positions were 
propagated to the SDSS epoch using proper motions from UCAC2 for
declinations below $41^\circ$. Because UCAC2 proper motions at
high declinations were not available, SDSS+USNO-B (Monet \etal\ 2003)
proper motions (Munn \etal\ 2004) were used for higher declinations.
In DR8, the UCAC2 proper 
motions in right ascension were incorrectly applied, introducing systematic
errors in right ascension of 5-10 mas.  For declinations above $41^\circ$,
the SDSS+USNO-B proper motions were not applied at all, introducing
systematic errors in both right ascension and
declination of typically 20-40 mas, and as high as 60 mas.

\item Previous SDSS data releases based the catalog right ascension and
declination values on the catalog {\tt objc\_rowc} and {\tt objc\_colc} frame
coordinates.  These coordinates use the $r$-band centroid for unsaturated stars
brighter than $r=22.5$, but for stars that are saturated in the $r$ filter but
unsaturated in another filter, or fainter than 22.5 in $r$ but better
exposed in another filter, uses the centroid from an optimal filter.
For DR8, the right ascension and declination values use the $r$-band centroid
for all stars.  This increases the statistical error for some stars
fainter than $r=22.5$ over that in earlier data releases.  For stars
saturated in $r$ but unsaturated in other filters, it can introduce systematic
errors of up to 100 mas compared to previous data releases.

\end{itemize}

The systematic errors introduced in DR8 are typically smaller than or comparable
to the 45 mas systematic errors that characterize the SDSS astrometry for
brighter stars.  Given these problems (which we plan to fix in a future data
release), we recommend that users interested in precise astrometry, especially
statistical studies of star positions at the $<0.1''$ level, use the
DR7 results.  For most applications, however, the quoted positions
should be acceptable.  Note in particular that the {\em proper motions}
tabulated in the CAS are only mildly affected by these problems, as
the systematic errors in position largely cancel when calculating
the proper motions.  The primary effects on the proper motions are to
introduce an additional systematic error with color of order 0.5
mas yr$^{-1}$, and to introduce an additional source of statistical error (in
right ascension only) for stars with $\delta > +41^\circ$ of order 1
mas yr$^{-1}$.

\subsection{Galaxy Zoo}
\label{sec:galaxyzoo}

Galaxy Zoo is a web-based project\footnote{\tt http://www.galaxyzoo.org} that used the collective efforts
of
hundreds of thousands of volunteers to produce morphological
classifications of galaxies. In the first phase of Galaxy Zoo, about
100,000 volunteers visually inspected $gri$ color composite images of
galaxies in the SDSS Main Galaxy spectroscopic sample (Strauss \etal\ 
2002) and classified them as ellipticals, spirals, mergers, or
star/don't know/artifact. In this phase, the project obtained 
more than $4\times10^7$ unique classifications. These basic
classifications are consistent with
those 
made by professional astronomers on sub-sets of SDSS galaxies (e.g.
they agree 90\% of the time with Fukugita \etal\ 2007), thus
demonstrating that the data provide a robust morphological catalog.
Full details on the classification process, including the operation
of
the site, are given in Lintott \etal\ (2008).

 The initial Galaxy Zoo data containing the basic classification data
 for 667,945 Main Galaxy sample galaxies (having measured redshifts in
 the range $0.001 < z < 0.25$ and clean $u$ and $r$ photometry in SDSS
 DR7) have recently been made public (Lintott \etal\ 2011). For each
 galaxy, this catalog includes weighted counts of volunteer ``votes''
 for the elliptical galaxy, spiral galaxy (split into clockwise or
 anticlockwise arms and edge-on/arms not visible), merger and
 ``star/don't know/artifact'' categories.  In addition, the catalog
 also includes votes corrected for perception bias effects and 
 information on confidence levels of the classification.  Those
 galaxies whose debiased votes give an unambiguous answer ($>80\%$) for their
 morphology are explicitly labeled as elliptical or spiral. Full details
 are given in Lintott \etal\ (2011). These initial Galaxy Zoo
 classifications are included in DR8, accessible through the Catalog
 Archive Server (\S\ref{sec:data}). The resulting catalog provides
 basic morphological classifications from visual inspection alone,
 providing an alternative to classifications based on parameters such
 as color, concentration, or structural parameters.  

\section{Spectroscopic Data}
\label{sec:spectro}

The principal changes in the spectroscopic data from those available
in DR7 are as follows: 
\begin{itemize} 
\item The inclusion of 211 new plates with spectroscopy of
  118,000 stars, from the SEGUE-2 survey (\S\ref{sec:segue2}). 
\item Improvements in the SEGUE Stellar Parameter Pipeline
  (SSPP; \S\ref{sec:sspp}). 
\item Improved data quality diagnostics on all plates
  (\S\ref{sec:spec_quality}). 
\item The release of 108 spectroscopic plates observed before summer
  2008 which were not included in DR7, and improved processing of a
  number of plates that targeted open and globular clusters used for SEGUE
  calibration (\S\ref{sec:new_plates}). 
\item Improved matching between the photometric and spectroscopic
  objects in the CAS (\S\ref{sec:matches}).
\end{itemize}
In addition, the redshifts and classifications included in DR8 are now
based on {\tt idlspec2d} instead of {\tt spectro1d}
(\S\ref{sec:specBS}), and we make available the results of an
independent code to measure galaxy emission line strengths and other
quantities derived from galaxy spectra
(\S\ref{sec:MPA}).

\subsection{SEGUE-2 Target Selection}
\label{sec:segue2}
  
The SEGUE-1 paper (Yanny \etal\ 2009) describes how that survey
selected spectroscopic targets, from extreme metal-poor star
candidates to low-mass stars to F-star tracers of the Galactic halo potential.
For SEGUE-2, these selection algorithms were refined in various ways,
as detailed in Rockosi \etal\ (2011; see also Eisenstein
\etal\ 2011).  We summarize the differences between SEGUE-1 and
SEGUE-2 here.  

In SEGUE-1, there were two pointings of 640 spectra on each 7 deg$^2$
plate area on the sky (hereafter, a ``tile''), one consisting of a
relatively short exposure on bright stars, and the other a longer
exposure on fainter stars.  The magnitude split between the bright and
faint plates was at $r = 17.8$ for $g-r < +0.55$ and $r = 17$ for $g-r
> +0.55$, allowing better S/N in the blue for cool stars.  The S/N for
stars as faint as $g=19.5$ was adequate to determine abundances using
the SSPP.  SEGUE-2 focused on spectroscopy of stars in the distant halo, 
and observed a single long-exposure pointing of 640 spectra on each tile,
allowing it to cover more sky in the year of the survey.  Fifty
percent of the stars with SEGUE-2 spectra have $17.4 < g < 18.9$, and
the median S/N per \AA\ of the SEGUE-2 spectra is 33.1.  For
comparison, 50\% of the SEGUE-1 spectra have $16.5 < g < 18.9$, and
the median S/N per \AA\ is 26.0.  A total of 211 observations were
made of 204 pointings in SEGUE-2, as shown in Figure~\ref{fig:sky_coverage}.

All the targets were selected using the SDSS imaging data and
recalibrated SDSS+USNO-B proper motions (Munn \etal\ 2004) from DR7.
Plates from Fall 2008 were designed using a preliminary version of the
DR7 data because the final version was not yet ready.  In order that
the survey target selection be reproducible, the photometry and
astrometry for all objects within the area of each plate, available at
the time the plates were designed, are included in a separate table in
the DR8 database.

SEGUE-2 increased the fraction of fibers devoted to candidate objects
in the outer halo over that in SEGUE-1, and modified the selection
criteria for red giant branch stars and blue horizontal branch stars
in order to increase the number of high-quality spectra for these
categories.  There were three target selection categories in SEGUE-1,
the F/G, G and dK,dM categories, which accounted for over half the
240,000 SEGUE-1 targets.  These were dominated by nearby main-sequence
stars, mostly in the disk, because they used only a simple color and
magnitude cut. Because SEGUE-2 observed about half the number of stars
per tile as SEGUE-1, we devoted only 100 fibers per plate to a
similar category called MS turnoff stars. The SEGUE-2 turnoff stars
are selected as targets with $18 < g < 19.5,+0.10 < g - r < +0.48$,
and range in distance from 6 to 13 kpc.

The SEGUE-2 selection of stars on the red giant branch (RGB) was
improved and extended to cooler giants based on the results from
SEGUE-1.  A total of 150 fibers per plate was devoted to this
category.  As in SEGUE-1, the selection required that the recalibrated
USNO-B+SDSS proper motions be consistent with 0 at $3\,\sigma$ to isolate
distant objects.  The confirmed low-gravity RGB stars from SEGUE-1, as
well as the globular and open cluster fiducial sequences from An
\etal\ (2008) and Clem, Vanden Berg \& Stetson (2008) were used to
identify regions of the $u-g$, $g-r$ color-color diagram where late K
and M giants are easily separated from the stellar locus.  The SEGUE-2
targeting also improved on the SEGUE-1 selection of warmer RGB stars
using the $l$-color (Lenz \etal\ 1998) indicator of
low-metallicity and (to a lesser extent) low-gravity stars.

The SEGUE-2 $ugr$ color selection of blue horizontal branch stars
includes only stars blueward of the old main sequence turnoff, $g-r <
+0.05$.  SEGUE-2 allocated as many as 100 fibers per plate to such
stars, but filled all those fibers only in the most crowded fields.
The fact that the density of blue horizontal branch stars and cool red
giant candidates was low was a major motivation to 
obtain only one tile per pointing and to maximize the area of
SEGUE-2.

New to SEGUE-2 are spectra of candidate old, metal-rich hypervelocity
stars using the color and proper motion selection criteria described in
Kollmeier \etal\ (2010).  In addition, 50 fibers per plate were
allocated to high velocity candidates with a $g-r$ color close to that
of the main sequence turnoff and velocities (based on proper motions)
estimated to be at least $3\,\sigma$ above 300 km s$^{-1}$. 
Finally, the selection of cool subdwarf and low-metallicity stars was
adjusted for improved efficiency based on the results of searches for
those objects using SEGUE-1 and SDSS spectra (L\'{e}pine \& Scholz
2008).

SEGUE-1 and SEGUE-2 spectroscopy was performed on only a small
fraction of the SDSS footprint, but both the SEGUE-1 and SEGUE-2 target
selection algorithms were applied to all the available imaging data; these
results are included in the DR8 database (\S\ref{sec:data}), as they may
be of use for statistical studies of the spatial 
distribution of various populations of stars.  

\subsection{Spectroscopic Classification and Redshift Measurement}
\label{sec:specBS}

The SDSS spectra are classified as stars, galaxies, or quasars, and
redshifts are determined with an automated routine.  As the DR6 paper
describes, this was done using two independent pipelines, one
({\tt spectro1d}) which worked by cross-correlation with a family of
templates, and emission-line fits, followed by eyeball inspection
of problematic cases, and another ({\tt idlspec2d} or {\tt specBS}) which
does direct $\chi^2$ fitting of templates to the spectra.  In DR8, we
only make the latter available; as described in the DR6 paper, the two
pipelines give substantially the same results for over 98\% of
spectra.  The \specBS\ pipeline has not been properly described in
print before, so we do so here.  

The classification and redshift-fitting procedures described below use
the spectrum and associated error estimate vectors (in the form of
inverse variances) to derive parameters of interest through
$\chi^2$ model-fitting to the spectra in pixel space (see Glazebrook,
Offer \& Deeley 1998 for an early version of this approach).  

A ``skymask'' is constructed and used to give zero weight in the fit
to pixels that show either bad sky subtraction in one of the 15-minute
exposures contributing to a given observation of a plate, or
extremely high relative sky brightness in all exposures.  The
condition of bad sky subtraction applies to all object spectra at a
given wavelength: we divide all sky-subtracted sky spectra on a given
plate by their associated error vectors, square these scaled values,
and mask wavelengths at which the 67th-percentile value of the
resulting quantity exceeds 3.  Bright sky is defined on an
object-by-object basis wherever the sky-line brightness exceeds the
sum of the extracted object flux plus ten times its associated error.
The skymask defined by these two conditions is grown by two extracted
pixels in either direction.  Regions of spectra affected by bad CCD
pixels or by excessive cosmic-ray hits are given an inverse variance
of zero by the two-dimensional extraction software routines, and are
not explicitly flagged in the redshifting and classification analysis.

Spectroscopic redshift
determination and object classification is done for all
  spectra without regard to the category by which they were targeted
  for spectroscopy, 
using four separate spectral template classes: galaxies, quasars,
stars, and cataclysmic variable stars.  

The galaxy class is defined by a rest-frame principal-component
analysis (PCA) of 480 main sample galaxies (Strauss \etal\ 2002) observed early in the SDSS, 
which is used to define a basis of four ``eigenspectra''
corresponding to the four most significant modes of variation in the
PCA analysis.  The redshifts of the galaxy PCA training sample are
established by fitting each spectrum with a linear combination of two
stellar template spectra and a set of narrow Gaussian profiles at the
wavelengths of common nebular emission lines.  The stellar template
spectra used in this procedure are obtained from the first two
components of a PCA analysis of ten velocity standard stars in M67
observed by SDSS (plate 321, observed on Modified Julian Date (MJD)
51612).  The galaxy PCA training sample redshifts 
were verified by visual inspection.

For all spectra, a range of trial galaxy redshifts is explored, from $z
= -0.01$ to $z = 1.00$ with a separation of 
138\,km\,s$^{-1}$ (i.e., two pixels in the reduced spectra).  At each
trial redshift, the galaxy eigenbasis is shifted accordingly, and the
error-weighted data spectrum is modeled as a minimum-$\chi^2$ linear
combination of the redshifted eigenspectra and a quadratic polynomial
to absorb low-order calibration uncertainties.  The $\chi^2$ value for
this trial redshift is stored, and the analysis proceeds to the next
trial redshift.  This procedure is facilitated by the
constant-velocity (constant log-wavelength) pixel width of the reduced
SDSS spectra, which permits redshifting of templates through simple
pixel shifting.  The trial redshifts corresponding to the five lowest
$\chi^2$ values are then re-determined locally to sub-pixel accuracy,
and errors in these values are determined from the curvature of the
$\chi^2$ curve at the position of the minimum.

Quasar redshifts are determined for all spectra in similar fashion to
the galaxy redshifts, but over a larger range of exploration ($z =
0.0333$ to $z = 7.00$) and with a larger initial velocity step
(276\,km\,s$^{-1}$).  The quasar eigenspectrum basis is defined by a
PCA of 412 quasar spectra with known redshifts, and an underlying
polynomial is allowed as well.  Star redshifts are
determined separately for each of 32 single sub-type templates
(excluding cataclysmic variables) using a single eigenspectrum plus a
cubic polynomial for each subtype, over a radial velocity range of
$\pm1200\rm \,km\,s^{-1}$.  Only the single best radial velocity is
retained for each stellar subtype.  Because of their intrinsic
emission-line diversity, cataclysmic variable stars are handled
differently from other stellar subtypes, with a three-component PCA
eigenbasis plus quadratic polynomial, over a radial velocity range of
${}\pm1000\rm \,km\,s^{-1}$.  Visual inspection of thousands of
galaxy, quasar, and cataclysmic variable star spectra (A. Bolton \& 
D. Schlegel 2011, private communication) demonstrate
that the eigenspectra modeling is adequate, in the sense that the
redshift error rate for spectra is of order 1\%, and the vast majority
of the failures are flagged with a redshift warning flag (see the
discussion in the DR6 paper).  

Once the best five galaxy redshifts, best five quasar redshifts, and
best stellar sub-type radial velocities for a given spectrum have been
determined, these identifications are sorted in order of increasing
reduced $\chi^2$, and the difference in reduced $\chi^2$ between each
fit and the next-best fit with a radial velocity difference of greater
than 1000\,km\,s$^{-1}$ is computed.  
The combination of redshift and template class that yields the lowest
reduced $\chi^2$ is adopted as the pipeline measurement of the
redshift and classification of the spectrum.  Redshifts are corrected
to the heliocentric frame.  Several warning flags can be set (Table 4
of the DR6 paper) to indicate low confidence in this
identification. The most common flag (``CHI2\_CLOSE'') is set
to indicate that the change in reduced $\chi^2$ between the best and
next-best redshift/classification is less than 0.01.

Stellar redshifts are recomputed using the ELODIE library spectra as
templates, after pruning to remove double and emission line stars and
anything else unsuitable for use as a velocity template.  These
redshifts represent our best estimate of the velocity of the star.  Note
however, that the velocity errors are
poorly characterized for the coolest (brown dwarf) and hottest (white
dwarf) stars.  See Schmidt \etal\ (2010) and West \etal\ (2011) for
independent radial velocity measurements of SDSS L and M dwarfs,
respectively. 

As described in the DR6 paper, there is a 
systematic offset of 7.3 km s$^{-1}$ in the stellar radial velocities
measured with the ELODIE templates;
this offset is corrected in the stellar parameters table in DR8.  The rms
plate-to-plate zero-point error in stellar velocities is
1.8 km s$^{-1}$, as measured using the approximately 30 stars that are repeated
on the bright and faint plates on each SEGUE-1 pointing.  At $r = 18$,
about the median S/N of the SEGUE stellar data, the total rms velocity
error (including the contribution from the zero point) is about 4.4
km s$^{-1}$, based on repeat observations. 

At the best galaxy redshift, the stellar velocity dispersion is also
determined by computing a PCA basis of eigenspectra
from the ELODIE stellar library (Prugniel \& Soubiran 2001), convolved
and binned to match the instrumental resolution and constant-velocity
pixel scale of the reduced SDSS spectra, and broadened by Gaussian
kernels of successively larger velocity width ranging from 0 to
$850\rm \,km\,s^{-1}$ in steps of $25\rm \,km\,s^{-1}$.  The broadened
stellar template sets are redshifted to the best-fit galaxy redshift,
and the spectrum is modeled as a least-squares linear combination of
the basis at each trial broadening, masking pixels at the position of
common emission lines in the galaxy-redshift rest frame.  The
dependence of $\chi^2$ on assumed velocity dispersion allows a
determination of the velocity dispersion and its error.  The error is
set to a negative value if the best value occurs at the high-velocity
end of the fitting range.  Reported best-fit velocity-dispersion
values less than about $100\rm \,km\,s^{-1}$ are below the resolution
limit of the SDSS spectrograph and are less reliable
(see the discussion in the DR6 paper).

Flux values, redshifts, line widths, and continuum levels are computed
for common rest-frame ultraviolet and optical emission lines by
fitting multiple Gaussian-plus-background models at their observed
positions within the spectra.  The initial-estimate emission-line
redshift is taken from the main redshift analysis, but is subsequently
re-fit non-linearly in the emission-line fitting routine.  All lines
are constrained to have the same redshift except for Lyman $\alpha$
(because of the bias induced by absorption from the Lyman $\alpha$
forest); note that this is not a perfect assumption for all quasar
lines (e.g., Richards \etal\ 2002b; Shen \etal\ 2008).
Intrinsic line widths are constrained to be the same for all emission
lines, with the exception of the hydrogen Balmer series, which is
given its own line width as a free parameter, and Lyman $\alpha$ and
NV~1241\AA, which each have their own free line-width parameters.  Known
3:1 line flux ratios for the [OIII]~4959,5007\AA\
and [NII]~6548,6583\AA\ doublets are imposed.  When the
signal-to-noise ratio of the line measurements permits doing so,
spectra classified as galaxies are sub-classified into AGN
and star-forming galaxies based upon measured [OIII]/H$\beta$ and
[NII]/H$\alpha$ line ratios (Baldwin, Phillips, \& Terlevich 1981,
hereafter BPT), and galaxies with very high equivalent
width in H$\alpha$ are sub-classified as starburst objects.  In the
following section, we describe an alternative method to measure emission-line
strengths.

\subsection{Quantities Derived from Galaxy Spectra}
\label{sec:MPA}

\subsubsection{Galaxy Emission Lines}
\label{sec:MPA_emline}
In measuring the nebular emission lines of galaxies, it is important
to properly account for the galaxy continuum which is very rich in
stellar absorption features.  The {\tt spectro1d} pipeline (Subbarao
\etal\ 2002) used in DR7 performs a simple estimate of the continuum
using a sliding median.  The {\tt idlspec2d} code described in
\S\ref{sec:specBS} uses a PCA technique to model the stellar
continuum, which has the disadvantage that it is not constrained to
produce astrophysically meaningful solutions.  In DR8 we offer a third
set of emission line measurements for galaxy spectra, which makes use
of stellar population synthesis models to accurately fit and subtract
the stellar continuum.  The code has been run on previous SDSS data
releases and the resulting measurements used for a variety of
scientific applications (e.g., Tremonti \etal\ 2004, Brinchmann \etal\
2004, Kauffmann \etal\ 2003b).  These data have been publicly
available\footnote{ \tt http://www.mpa-garching.mpg.de/SDSS/} since
DR4; we are making them accessible through the SDSS data release for
the first time with DR8.  We refer to this set of line measurements as
the MPA-JHU measurements, after the Max Planck Institute for Astrophysics
and the Johns Hopkins University where the technique was developed.  We provide MPA measurements for
all objects that {\tt idlspec2d} calls a galaxy; see
\S\ref{sec:specBS}.  We briefly describe the technique here; details
can be found in Tremonti \etal\ (2011).

We first scale each galaxy spectrum to match its $r$-band fiber
magnitude, and 
correct each spectrum for Galactic extinction following SFD
and the O'Donnell (1994) attenuation curve. 
We adopt the basic assumption that any galaxy star formation history
can be approximated as a sum of discrete bursts.  Our library of
template spectra is composed of single stellar population models
generated using the population synthesis code of
Bruzual \& Charlot (2003).  We have used a new version kindly made
available by the authors which incorporates the MILES empirical
spectral library (S\'anchez-Bl\'azquez \etal\ 2006; these spectra cover
the range 3525-7500 \AA\ with 2.3 \AA\ FWHM). The spectral-type and metallicity
coverage, flux-calibration accuracy, and number of stars in the
library represent a substantial improvement over previous
libraries. Our templates include models of ten different ages (0.005,
0.025, 0.1, 0.2, 0.6, 0.9, 1.4, 2.5, 5, 10 Gyr) and four metallicities
(1/4, 1/2, 1, 2.4 $Z_{\sun}$).  For each galaxy we transform the
templates to the measured redshift and velocity dispersion and
resample them to match the data.  To construct the best-fitting model
we perform a non-negative least squares fit to a linear combination of
our ten single-age populations, with internal dust attenuation
modeled as an additional free parameter following 
Charlot \& Fall (2000).  Given the S/N of the spectra, we model 
galaxies as single 
metallicity populations and select the metallicity that yields the
minimum $\chi^2$.

After subtracting the best-fitting stellar population model of the
continuum, we remove any remaining residuals (usually of order a few 
percent) with a sliding 150-pixel median, and fit all the nebular emission
lines simultaneously as Gaussians.  In doing so, we require that the
Balmer lines (H$\delta$, H$\gamma$, H$\beta$, and H$\alpha$) have the
same line width and velocity offset, and likewise for the forbidden lines
(e.g., [\ion{O}{2}]~$\lambda\lambda 3726, 3729$, 
[\ion{O}{3}]~$\lambda\lambda4959, 5007$,
[\ion{N}{2}] $\lambda\lambda6548, 6584$, 
[\ion{S}{2}]~$\lambda\lambda6717, 6731$).  
We take into account the
wavelength-dependent instrumental resolution of each fiber, which is
measured by the {\tt idlspec2d}
pipeline from the arc lamp images.  

\begin{figure}[t]\centering\includegraphics[width=10cm]
{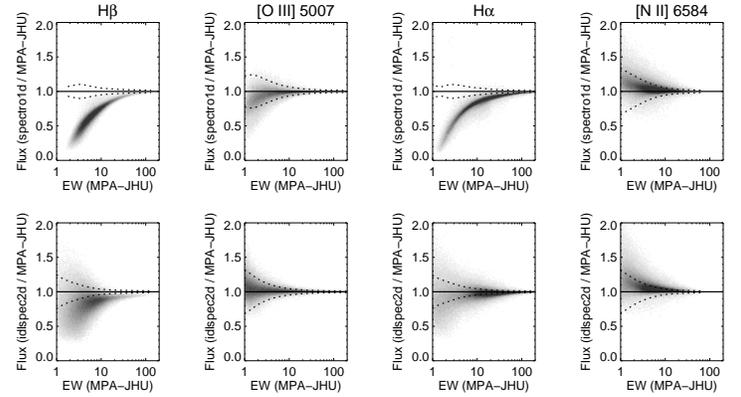}
\caption{Ratio of the {\tt spectro1d} and {\tt idlspec2d} emission
  line flux measurements with those of the MPA-JHU pipeline, as a
  function of rest-frame equivalent width, for galaxies in DR8 with
  emission line measurements with greater than $3\sigma$ significance.
  In performing this comparison, we have put all measurements on a
  common scale by removing the Milky Way reddening and and
  spectrophotometric zeropoint corrections from the MPA-JHU line
  measurements. The remaining differences are due to the different
  methods of modeling the stellar continuum. The dotted lines show the
  deviation expected due to random error.
\label{fig:mpa_fnal_comp}}
\end{figure}

In Figure \ref{fig:mpa_fnal_comp} we explore the differences in the
line fluxes measured by the MPA-JHU, {\tt spectro1d} and {\tt idlspec2d}
codes resulting from the differences in modeling the stellar
continuum.  The line fluxes of [O III]~$\lambda5007$ and 
[NII]~$\lambda6584$ are generally consistent within the errors.  
The Balmer lines are systematically underestimated by {\tt spectro1d} at
low equivalent widths because stellar Balmer absorption has not been 
accounted for by the smooth continuum model they used.  The differences 
are smaller when comparing the
MPA-JHU and {\tt idlspec2d} measurements, since both codes 
model the stellar continuum in detail, but they are still significant for
H$\beta$.   
 
The {\tt idlspec2d} and MPA-JHU codes also show significant differences in
equivalent width measurements of Balmer lines.  The {\tt idlspec2d}
code records the 
continuum at line center of the best fit stellar continuum model
(corresponding to the trough of Balmer stellar absorption lines),
while the MPA-JHU code median smooths the emission line
subtracted spectrum by 100 pixels ($\sim6900$ km~s$^{-1}$) before
recording the continuum at line center.  
For
galaxies with significant intermediate age stellar populations, the
differences between the two continuum measurements can be as large as
30\%, which has a correspondingly large effect on line equivalent
widths. 

\subsubsection{Physical Properties of Galaxies}
DR8 also includes a number of galaxy physical parameters 
derived by the MPA-JHU group available: 

\begin{itemize}

\item {\bf BPT classification:} We supply emission line
  classifications based on the BPT diagram, [N~II]~6584/H$\alpha$ vs. [O~III]~5007/H$\beta$.
  Galaxies are divided into {\it Star Forming}, {\it Composite}, {\it
  AGN}, {\it Low S/N Star Forming}, {\it Low S/N AGN}, and {\it
 Unclassifiable} categories as outlined in Brinchmann \etal\ (2004).

\item {\bf Stellar Mass:} Stellar masses are calculated using the
  Bayesian methodology and model grids described in Kauffmann \etal\
  (2003a).  The spectra are measured through a $3''$ aperture, and
  therefore do not represent the entire galaxy.  We therefore base our
  model on the $ugriz$ galaxy 
  photometry alone (rather than the spectral indices D$_n$(4000) and
  H$\delta_A$ used by Kauffmann \etal\ 2003a). We have
  corrected the photometry for the small contribution due to nebular
  emission using the spectra.  We estimate the stellar mass within the
  SDSS spectroscopic fiber aperture using fiber magnitudes and the total stellar
  mass using model magnitudes.  A Kroupa (2001) initial mass function
  is assumed.  We output the stellar mass corresponding to
  the median and 2.5\%, 16\%, 84\%, 97.5\% of the probability
  distribution function.

\item {\bf Nebular Oxygen Abundance:} Nebular oxygen abundances 
  are estimated from the strong optical emission lines 
  ([O II]~3727, H$\beta$, [OIII]~5007, 
  [NII]~6548,~6584 and [SII]~6717, 6731) using the Bayesian
  methodology outlined in Tremonti \etal\ (2004) and Brinchmann \etal\ (2004).
  Oxygen abundances are only computed for objects classified as
  {\it Star Forming}. We output the value of $\rm 12+\log(O/H)$ at the median 
  and 2.5\%, 16\%, 84\%, 97.5\% of the probability distribution 
  function.  

\item {\bf Star Formation Rate:} Star formation rates (SFRs) are computed
  within the galaxy fiber aperture using the nebular emission lines 
  as described in Brinchmann \etal\ (2004). SFRs outside of the fiber are
  estimated using from fits of model grids to the $u,g,r,i,z$
  photometry outside the fiber, following the method described in
  Salim \etal\ (2007)\footnote{The SFRs provided on the
    MPA-JHU website use a slightly  different technique for galaxies
    for weak emission lines, as will be described in Brinchmann \etal\
    (in preparation).}. The same technique was also applied to
  estimate 
  SFRs in AGN and galaxies with weak emission lines. We report both the fiber SFR and the total SFR
  at the median and 2.5\%, 16\%, 84\%, 97.5\% of the probability
  distribution function.

\item {\bf Specific SFR:}  The Specific SFR (the ratio SFR to the
  stellar mass)
 has been calculated by combining the SFR and stellar mass likelihood
 distributions as outlined in Appendix A of Brinchmann \etal\ (2004). 
 We report both the fiber and the total specific SFR
 at the median and 2.5\%, 16\%, 84\%, 97.5\% of the probability
 distribution function.

\end{itemize}

\subsection{Changes to SSPP}
\label{sec:sspp}

The SEGUE Stellar Parameters Pipeline (SSPP; Lee \etal\ 2008ab,
Allende Prieto \etal\ 2008) fits models to SDSS
spectra of stars in order to determine surface temperature, gravity,
and metallicity.  
The pipeline was refined for SEGUE-2 to improve the
parameter estimates, as described in the Appendix of Smolinski \etal\
(2011).  This refined version, which we summarize here, has 
been used for the DR8 processing. 

The SSPP uses multiple techniques to estimate [Fe/H], 
effective temperature and surface gravity. Each of these methods is considered valid over
a particular range of $g-r$ and S/N, and some methods are more
accurate or better calibrated at low or high metallicity.   To choose
between them, we compare the observed and model spectra at the 
metallicity given by each method, and reject those for which the
correlation coefficient between the spectra or the mean residuals are
poor.  
This approach has improved the accuracy of metallicity estimates for
stars up to solar metallicity, as demonstrated in particular by the
SSPP parameters for 
stars in M67 in Smolinski \etal\ (2011).   Further work on reducing bias in the
SSPP in other parts of the HR diagram came from adjusting the $g-r$ and S/N
ranges for some estimators, and recalibration of others using the
cluster plates (\S\ref{sec:new_plates}) and high-resolution data taken
on other telescopes.  The SSPP reports stellar
parameters for stars in the range $-0.3 < g-r < 1.3$, but below $g-r =
0.0$ ($T_{eff}=7500$ K) or above $g-r=0.8$ ($T_{eff}=4500$ K), the
errors in $T_{eff}$ and $\log g$ become appreciably larger.

The SSPP now also includes estimates of metallicity, gravity and
temperature based on the spectra alone, with no photometric
information.  These ``spectroscopic only'' parameter estimates are
more reliable in regions of high extinction 
(Cheng \etal\ 2011).  Finally, the SSPP reports metallicity and
gravity estimates made with the effective temperature determined from
a color-temperature relation; these may provide more reliable
parameter estimates for low-metallicity stars.

\subsection{Spectroscopic Data Quality}
\label{sec:spec_quality}

Each spectroscopic plate is assigned a quality ({\tt PLATEQUALITY}) with one of
three values: ``good,'' a good science quality plate; ``marginal,'' an
acceptable plate, but lower quality than good plates; and ``bad,'' a
plate with results that should be treated with skepticism. 

The {\tt PLATEQUALITY} value is set independently for each observation
(labeled by the Modified Julian Date of the observation) of each
plate.  
For Legacy plates, the definition of plate quality is based on the
median squared signal-to-noise ratio per spectroscopic pixel for targets at
$g_{\mathrm{fiber}}=20$ 
($(S/N)^2$ in what follows) and the fraction $f_{\rm bad}$ of pixels in
the sky fibers that have $\chi^2>4$ in the model for the sky spectrum
in any of the contributing exposures.  In particular, a plate with
$(S/N)^2>15$ and $f_{\mathrm{bad}}< 0.05$ is deemed ``good''; a plate
with $(S/N)^2>9$ and $f_{\mathrm{bad}}< 0.13$ is deemed ``marginal'';
and otherwise it is deemed ``bad.''

For SEGUE plates, the conditions are based on the signal-to-noise
ratio of main-sequence turnoff stars at $g=18$. For faint SEGUE-1
plates, a plate with $(S/N)^2>16$ is deemed ``good.'' For bright
SEGUE-1 plates, a plate with $(S/N)^2>7.5$ is deemed ``good.''  
SEGUE-2 plates with $(S/N)^2>10$  are considered ``good''. SEGUE-1 and SEGUE-2 plates do
not have a ``marginal'' quality designation.  Finally, for
plates observed during the first stages of commissioning, low Galactic
latitude plates, and cluster plates (\S~\ref{sec:new_plates}), the quality is set by
visual inspection of the data. 

Three additional flags provide more detail on the nature of the
plate. {\tt IS\_BEST} is set to 1 if a given observation is the best
observation of a plate (whether or not it is marked as bad), and 0
otherwise.  {\tt IS\_PRIMARY} is set to 1 if the plate is the best
observation of a given plate (i.e., {\tt IS\_BEST} is set), {\it
  and} the observation is not marked as ``bad,'' and 0 otherwise.
Finally, {\tt IS\_TILE} is set to 1 if the plate is the best Legacy
plate covering its location, and 0 otherwise; the definition of the
Legacy spectroscopy is the union of all plates with {\tt IS\_TILE}
set.  A plate can only be {\tt IS\_TILE} if it is also {\tt
  IS\_PRIMARY}.

Selecting plates which are not ``bad'' will yield a good sample of
spectra.  Nevertheless, many of the ``bad'' plates actually contain
useful data (in particular, many highly reliable redshifts). However,
bad plates should be treated with care (in particular, they may have
poor spectrophotometry or residual sky subtraction problems).

\subsection{New and Reprocessed Plates}
\label{sec:new_plates}

In DR7 and previous data releases, there were a number of observations of plates
that had been observed and reduced, but not included in the releases
because they were of lower quality and/or were repeats of other
plates. In DR8, we are releasing 108 such plates, with improved
quality flags so that marginal or bad plates can be flagged in
analysis.  Twelve of these 108 plates are new, in the sense that they
are not simply repeats of observations already included in DR7.  Of
these 108 plates, 24 are classified ``good''.

SEGUE observed stars in a number of well-studied
open and globular clusters, including 
M92, NGC5053, M53, M15, M13, M2, M3, NGC 2420, M67, NGC 6791, M71, Be
29, M35, NGC 2158 and NGC 7789 (Rockosi \etal\ 2011, Smolinski \etal\
2011, Ma \etal\ 2011).  These clusters
have well-measured metallicities and allow us to sample regions of the
HR diagram that we do not otherwise probe in the SDSS, so observations
of these clusters are invaluable for calibrating the outputs of the
SSPP.  These so-called 
``cluster plates'' were made available in DR7, but we faced some
challenges in reducing them. 
Difficulties
included background contamination in the target, flux-calibration and
sky fibers due to the crowded fields, the lack of good-quality
reductions of the relevant SDSS photometric data (indeed, we did not
have SDSS imaging data at all for some clusters), and the large range
of brightnesses of targets on a single plate, giving rise to
cross-talk between adjacent fibers.  For DR8 the cluster plates
were reprocessed using careful iterative selection of the sky and
flux-standard fibers.  The required changes in the reduction procedure
were small enough that the
goal of having uniform reductions for the cluster calibration stars
and the survey plates was met.

Because of the difficulty in finding good photometric standards for
the reductions of the cluster plates, there are some low-level, large-scale
residuals in the spectrophotometric solution.  These residuals are corrected in
the continuum normalization procedure in the SSPP, and the SSPP
parameters are unaffected.  However, users of these spectra should be
aware of these and other possible systematic errors in the flux calibration.

\subsection{Matching Photometry to Spectroscopy}
\label{sec:matches}

In DR8, we introduce a new method for matching the photometry to the
spectroscopy. Instead of a purely positional match that searches for
the nearest photometric object center to a spectrum, we search for the
object that, according to the photometric reductions, contributes the
greatest amount of light to the spectrum.  In detail, we quantify the
contribution of light using a $3''$ diameter aperture in the
$r$-band. While this ``flux-based'' match is the default that we
provide in the data release, the ``position-based'' match is also
provided.  We do not correct for proper motion of stars between the
time that the images and the spectra were taken.  

The ``flux-based'' match is usually appropriate and typically more
accurate for large, nearby galaxies. In particular, the latest
photometric pipeline version often deblends parent objects into children
differently than the version that was used for targeting. Therefore, the
spectrum of a galaxy might be significantly offset
from the location that we now deem to be its ``center.'' The ``flux-based''
matches recover many such cases. The ``position-based'' match is
important for other purposes such as spectrophotometry.

In more detail, we first execute a purely positional match to the
primary photometric catalog for each spectrum, using a $2''$
matching criterion. For each spectrum, the matching photometric object
id is stored in the field {\tt ORIGOBJID} in the files and in the database.
For the $\sim 1\%$ of spectra that have no position-based match, we
find the primary imaging field that contains the location of the
spectrum. If there are no detected pixels at the location of the
spectrum (that is, if it is not contained in a ``parent'' object) then
the object is unmatched. This happens for about 90\% of the objects
without a position-based match; these objects are typically sky fibers
or transient objects such as satellites, in cases where the
primary imaging field in the final photometric catalog differs from
the original field used to target the spectroscopy.

Some spectra with no position-based match nevertheless fall within the
boundaries of some ``parent'' object.  In these cases, we perform
$3''$ diameter aperture photometry in the $r$-band at the location of
the spectrum, using the atlas images of the parent and all of its
children. The flux-based match is designated to be the child
that contributes the most flux to the parent, and we store its object
id as the {\tt BESTOBJID} associated with the spectrum.

Finally, for spectra with a position-based match, we compare the $3''$
fiber flux with a $3''$ aperture flux based on the radial profile
measured by {\tt photo}.  The fiber magnitude is based on the parent
atlas image, whereas the radial profile is calculated using only the
child atlas image. Therefore, in cases where our aperture flux is less
than 50\% of the fiber flux, the light in the fiber is dominated by
other objects.  In those cases, we perform aperture photometry at the
fiber location on the atlas images of the parent and all children. We
select the child with the most flux as the flux-based match, and store
its object id as the {\tt BESTOBJID} associated with the spectrum.

About 0.5\% of all spectra have flux-based matches that differ from
the position-based matches. Typically, half of these are cases where
the 
photometry is irretrievably bad in some way (such as a long
satellite trail or airplane). The other half are cases where 
the flux-based match appears more appropriate when one examines the
images by eye; that is, where the
redshift of the spectrum should be associated with the flux-based
match in the photometric catalog.

\section{Data Distribution}
\label{sec:data}

In SDSS-I/II, the data were distributed with two different portals.
The Catalog Archive Server (CAS) is a database containing catalogs of
SDSS objects (both photometric and astrometric) that allowed 
queries on their measured attributes.  The Data Archive Server (DAS)
consists of flat files containing the images themselves, the
catalogs, the spectra, and other data products.  We continue to use the CAS for
DR8\footnote{{\tt http://skyserver.sdss3.org/dr8/}}; it is largely
unchanged, although some obsolete tables and schema have been removed.

The design of the DR7 CAS considered the SDSS Legacy survey to be
fundamental.  Thus imaging objects that fell outside the Legacy
footprint were flagged as secondary.  The DR8 CAS does not keep this
distinction; it treats all imaging runs as equivalent and uses the
uniform results from the resolve algorithm (\S\ref{sec:resolve})
across the entire unique imaging area.

The DAS functionality has been replaced with the SDSS-III Science Archive
Server (SAS)\footnote{{\tt http://data.sdss3.org}}, which has a
similar, but not identical, directory structure.  In SDSS-I/II, the
names of various fields and attributes differed between the DAS and
CAS.  More importantly, there was not a perfect match between the
contents of the two: for example, there were imaging runs and
spectroscopic plates available in the DAS that were not present in the
CAS.  We have endeavored to couple the CAS and the SAS more closely in
DR8.  To 
a very good approximation, the data contained in the two are the same,
though packaged differently.  In particular, unlike DR7, {\em all} the
normal imaging scans included in the SAS are in the CAS as well.

In the DR7 DAS, the photometrically and astrometrically calibrated versions of
these files were called {\tt tsObj} or {\tt drObj} files; the
nomenclature of the uncalibrated and corresponding calibrated
quantities was not always consistent (for example, some calibrated
quantities had names, like {\tt psfCounts}, that erroneously implied
that they were not calibrated).  This situation has been rectified in the
so-called {\tt photoObj} files found in the SAS and in the tables in
the CAS.  Similarly, the metadata files
describing each field, which in DR7 were called {\tt tsField} files,
have a changed format, called {\tt photoField} files, which includes information about the
ubercalibration. The full data
model with a definition of all terms
may be found on the DR8 website.

In addition to the {\tt photoObj} files, we also provide a much more
compact version of the catalog called the ``datasweeps,'' in the {\tt
  calibObj} files.  These files mirror the {\tt photoObj} files but
only list the most commonly used attributes for each object, and only
retain objects with a reasonable detection\footnote{Defined to be
  those stars for which the PSF magnitude in at least one of
  $(u,g,r,i,z)$ is brighter than $(22.5, 22.5, 22.5, 22, 21.5)$, and
  those galaxies for which one model magnitude is brighter than $(21,
  22, 22, 20.5, 20.1)$, after correction for Galactic extinction
  following SFD. This criterion
  excludes roughly 23\% of the objects.}
   in at least one band. The
datasweeps are convenient for users who need basic information for
all objects in a compact form.

In DR7, only calibrated asinh magnitudes (Lupton \etal\ 1999) were
tabulated, with names like {\tt psfMag}.  With DR8, we also include,
for all photometric quantities, the linear flux density (i.e., no
logarithms or asinh!), in units of ``nanomaggies'' (Finkbeiner \etal\
2004), with names like {\tt psfFlux}.  A nanomaggie (nMgy) is defined
as the flux density (per unit frequency) of a 22.5 AB magnitude
object, in any band.  Given the definition of AB magnitudes (Oke \&
Gunn 1983),
$$1\, \rm nMgy = 3.631 \times 10^{-6}\,Jansky = 3.631 \times
10^{-29}\,erg\,s^{-1}\,cm^{-2}\,Hz^{-1}.$$ 

As in DR7, SAS makes available corrected frames of each field, in
which defects have been interpolated over.  However, unlike DR7, the DR8 versions of
these files contain flux values calibrated in nanomaggies, have a
global best-fit sky model (\S\ref{sec:globalsky}) subtracted, and
have a proper WCS header. The calibration
and sky-subtraction information is bundled with the files and can be
easily backed-out if necessary.

Finally, the SAS user interface is quite different from that of the
DR7 DAS. In
addition to allowing for searches for spectra based on coordinates,
redshifts, target flags, and fiber identification numbers, it provides
an interactive interface to  plot the spectra.  It also allows
coordinate searches for fields, as well as returning FITS mosaics that
stitch together overlapping fields.

\section{Conclusions} 
\label{sec:summary}

This paper describes the eighth data release of the Sloan Digital
Sky Survey, consisting of all the SDSS data taken through
Summer 2009, together with the final imaging of the Southern Galactic
Cap completed in 2010 January.  The images cover a footprint of over
14,500 deg$^2$; including repeat observations, the total quantity of
imaging data is more than twice this value. All these data have been reprocessed
with an updated version of the photometric pipeline, which gives
modest improvements to the photometry of bright galaxies and fainter
galaxies near them.   In addition, DR8 contains the spectra of over
1.6 million galaxies, quasars, and stars, including 118,000 new
stellar spectra from the SEGUE-2 survey, as well as 108 plates of data
not previously released. 

With the completion of the imaging survey, the SDSS camera has been
retired.  SDSS-III is described in detail in
Eisenstein \etal\ (2011); it will continue through 2014. This release
contains data from two of its four surveys: SEGUE-2, and the
imaging component of BOSS.  BOSS spectroscopy has started, and its 
first year of data will be made available as part of the ninth data
release.  Plots showing the quality of those data may be found in
Eisenstein \etal\ (2011) and White \etal\ (2011).  In addition, the
MARVELS survey is well underway, and the first scientific results have
been published (Lee \etal\ 2011).  Finally, APOGEE will probably have seen
first light by the time this article is published, and data
from that survey will first be released publicly in the tenth data
release.

We thank the referee, Andrew West, for comments that improved the
paper. 
Funding for SDSS-III has been provided by the Alfred P. Sloan
Foundation, the Participating Institutions, the National Science
Foundation, and the U.S. Department of Energy. The SDSS-III web site
is http://www.sdss3.org/.

SDSS-III is managed by the Astrophysical Research Consortium for the
Participating Institutions of the SDSS-III Collaboration including the
University of Arizona, the Brazilian Participation Group, Brookhaven
National Laboratory, University of Cambridge, University of Florida,
the French Participation Group, the German Participation Group, the
Instituto de Astrofisica de Canarias, the Michigan State/Notre
Dame/JINA Participation Group, Johns Hopkins University, Lawrence
Berkeley National Laboratory, Max Planck Institute for Astrophysics,
New Mexico State University, New York University, Ohio State
University, Pennsylvania State University, University of Portsmouth,
Princeton University,  the Spanish Participation Group,
University of Tokyo, University of Utah, Vanderbilt University,
University of Virginia, University of Washington, and Yale
University.

\end{document}